\documentclass[11pt]{article}
\usepackage{array}
\usepackage{amsmath}
\usepackage{amssymb}
\usepackage{graphicx}
\usepackage{amsthm}
\usepackage{latexsym}
\newtheorem{theorem}{Theorem}
\newtheorem{cor}{Corollary}

\newtheorem{prop}{Proposition}
\newtheorem{lemma}{Lemma}
\newtheorem{example}{Example}

\newtheorem{remark}{Remark}
\newtheorem{algorithm}{Algorithm}

\newcommand{\beq}{\begin{equation}}
\newcommand{\eeq}{\end{equation}}
\newcommand{\barr}{\left[\begin{array}}
\newcommand{\earr}{\end{array}\right]}

\newcommand{\bpf}{\emph{Proof}\/:}
\newcommand{\epf}{\hfill$\Box$}

\newcommand{\ff}{\ensuremath{\mathbb{F}}}

\newcommand{\bi}{\begin{itemize}}
\newcommand{\ei}{\end{itemize}}
\newcommand{\bnum}{\begin{enumerate}}
\newcommand{\enum}{\end{enumerate}}
\newcommand{\bc}{\begin{center}}

\newcommand{\supp}{\mbox{supp}}
\begin{document}
\title{An algorithm for Boolean satisfiability based on generalized orthonormal expansion}
\author{Virendra Sule\\Department\ of Electrical Engineering\\Indian Institute of Technology Bombay, India\\vrs@ee.iitb.ac.in}
\maketitle

\begin{abstract}
This paper proposes an algorithm for deciding consistency of systems of Boolean equations in several variables with co-efficients in the two element Boolean algebra $B_{0}=\{0,1\}$ and find all satisfying assignments. The algorithm is based on the application of a well known generalized Boole-Shannon orthonormal (ON) expansion of Boolean functions. A necessary and sufficient consistency condition for a special class of functions was developed in \cite{sule} using such an expansion. Paper \cite{sule} develops a condition for consistency of the equation $f(X)=0$ for the special classes of Boolean functions 1) $f$ in $B(\Phi(X))$ for an ON set $\Phi$ of Boolean functions in $X$ over a general Boolean algebra $B$ and 2) $f$ in $B(X_{2})(\Phi(X_{1}))$. The present paper addresses the problem of obtaining the consistency conditions for arbitrary Boolean functions in $B_{0}(X)$. Next, the consistency for a single equation is shown equivalent to another system of Boolean equations which involves the ON functions and characterizes all solutions. This result is then extended for Boolean systems in several variables over the algebra $B_{0}=\{0,1\}$ which does not convert the system into a single equation. This condition leads to the algorithm for computing all solutions of the Boolean system without using analogous resolution and determine satisfiability. For special systems defined by CNF formulas this algorithm results into an extension of the DPLL algorithm in which the \emph{splitting rule} is generalized to several variables in terms of ON terms in the sense that splitting of CNF set in a single variable $x$ is equivalent to ON terms $x,x'$. 
\end{abstract}

\noindent
Category: cs.CC, cs.SC, ms.RA\\
ACM class: I.1.2, F.2.2, G.2\\
MSC class: 03G05, 06E30, 94C10.

\section{Introduction}
The problem of Boolean satisfiability over the Boolean algebra $B_{0}=\{0,1\}$ is defined by a system of equations
\begin{equation}\label{boolsystem}
f_{i}(X)=g_{i}(X), i=1,2,\ldots,N
\end{equation}
where $f_{i}(X),g_{i}(X)$ are Boolean functions of $n$ Boolean variables $x_{1},\ldots,x_{n}$ denoted $X$ with $B_{0}$ co-efficients where $B_{0}=\{0,1\}$ is the well known two element Boolean algebra. Boolean functions in $n$ variables $X$ with co-efficients in a general Boolean algebra $B$ are formal expressions with Boolean operations $+,.,'$ between variables, expressions and constants in $B$. Equivalence classes of these expressions, having same $B$ values when variables are assigned from $B$ are called Boolean functions. These form a Boolean algebra which we shall denote alternatively by $B(X)$ and also by $B(n)$. For definition and properties of different types of Boolean functions we refer \cite{rude,brow}. The system (\ref{boolsystem}) is said to be \emph{consistent} or \emph{satisfiable} if there exist values (called \emph{assignments}) of variables $X$ in $B_{0}^{n}$ at which the equations hold true. 

In order to introduce the problem addressed in this paper we need to consider previous results and background. A set $\Phi=\{\phi_{1},\ldots,\phi_{m}\}$ in a Boolean algebra $B$ is said to be \emph{orthogonal} (OG) of \emph{order} $m$ if $\phi_{i}\phi_{j}=0$ for $i\neq j$ and is called \emph{orthonormal} (ON) if in addition to this condition the members satisfy
\[
\sum_{i=1}^{m}\phi_{i}=1
\]
Such a set is said to be \emph{reduced} if none of the members $\phi_{i}$ are zero in $B$. In this paper by an ON set we shall always refer to a reduced ON set. For a background of ON systems in general Boolean algebra the reader is referred to \cite{rude}. If $\Phi$ is an ON set of order $m$ in the algebra of Boolean functions $B(X)$, by the well known result \cite[proposition 3.14.1]{brow} a Boolean function $f(X)$ in $B(X)$ has an expansion as
\begin{equation}\label{ONexp}
f(X)=\sum_{\phi_{i}\in\Phi}\alpha_{i}(X)\phi_{i}(X)
\end{equation}
This expansion generalizes the well known Shannon expansion (proposed to be called Boole-Shannon expansion in \cite{sule}). The expansion co-efficient functions $\alpha_{i}(X)$ for a given ON set $\Phi$ are not unique as for every $i$ any function in the range $[f\phi_{i},f+\phi_{i}']$ satisfies the expression.

\subsection{Previous results}
Given an ON set $\Phi$ in $B(X)$ the set $B(\Phi)$ is defined in \cite{sule} by
\[
B(\Phi)=\{\sum_{\phi_{i}\in\Phi}\alpha_{i}\phi_{i},\alpha_{i}\in B\}
\]
The problem addressed in \cite{sule} is to express the condition for consistency of a single Boolean equation $f(X)=0$ for $f$ in $B(\Phi)$ in terms of the expansion co-efficients $\alpha_{i}$. It is shown in \cite[Theorem 2]{sule} that $f(X)=0$ for such a function is consistent iff
\[
\prod_{i}\alpha_{i}=0
\]
in $B$. Implications of this result in terms of elimination of variables is discussed in \cite{sule} and leads to a procedure for deciding consistency by eliminating partial set of variables successively. If the function $f$ belongs to $B(X)$ in $n$ variables $X$ while $\Phi$ are defined over a subset $X_{1}$ in a partition $X=\{X_{1},X_{2}\}$ then if $f$ belongs to $B(X_{2})(\Phi(X_{1}))$ the expansion co-efficients $\alpha_{i}$ exist in $B(X_{2})$. Under these conditions it is further shown in \cite[Corollary 1]{sule} that $f=0$ is consistent iff 
\begin{equation}\label{product}
\prod_{i}\alpha_{i}(X_{2})=0
\end{equation}
is consistent. This way the number of variables can be eliminated successively to finally arrive at a condition for consistency of $f=0$. The product of co-efficients above is analogous to the well known resolution of clauses in CNF SAT studies \cite{hand}.

\subsection{Problems addressed}
The aim of this paper is twofold. First we want to determine the consistency conditions for $f(X)=0$ w.r.t. an ON set $\Phi$ in $B(X)$ for general functions (not restricted to $B(\Phi)$). Due to this generality an ON expansion of a function $f(X)$ w.r.t. $\Phi$ may have the co-efficients $\alpha_{i}(X)$ as functions which may not be constants in $B$. Moreover as a functions in $B_{0}(X)$ the arguments of co-efficients and ON functions might also overlap. Previous proof of consistency \cite{sule} considers the case when only the constant co-efficients $\alpha$ are involved or the when the arguments of all co-efficients $\alpha_{i}$ in the expansion are distinct from those of the ON set $\Phi$. Hence this proof is not applicable in the present case. 

Further, the aim of this paper is to develop an algorithm for deciding Boolean satisfiability (and to determine all solutions) of Boolean \emph{systems} (\ref{boolsystem}) where functions in the system belong to $B_{0}(X)$. Determining all solutions of a Boolean system is interpreted in the sense that the original Boolean system is decomposed in terms of multiple systems in smaller number of variables whose union is a projection of solution set of the original system. Hence the original system is consistent iff at least one of the smaller systems is consistent. This decomposition can be iteratively applied to finally get in principle smallest Boolean systems involving no variables hence of the form $a=b$ for $a,b$ in $B_{0}$ which decide consistency. While iteratively decomposing the original Boolean system, several trivial assignments of variables can be discovered (such as those involving unit clauses $x=1$ or $x'=1$ as well as pure literals as in CNF satisfiability by DPLL algorithm). Such a decomposition strategy for Boolean systems is desirable for efficient computation of satisfiability and solutions.

\subsubsection{Computational aspects}
To explain the computational aspects we first note that algorithms for Boolean satisfiability by elimination involve two impediments
\begin{enumerate}
\item The consistency condition or the process of elimination of variables involves computation of eliminant of a single equation $f(X)=0$. Hence it is necessary to convert a system of Boolean equations (\ref{boolsystem}) into such an equation by constructing the single associated function
\[
F(X)=\sum_{i=1}^{N}(f_{i}(X)\oplus g_{i}(X))
\]
Hence an algorithm for consistency of the system has to handle an overload of computing $F(X)$ which may be heavy in memory requirement. 
\item The resolution step involves computation of the product (\ref{product}) whose consistency needs to be determined. Computation of this product is another (memory) overhead of such algorithms. 
\end{enumerate} 

Due to these difficulties elimination of variables for computation may be advisable only when number of variables to be eliminated is small enough relative to memory available. Hence it is desirable for an algorithm for satisfiability of a system (\ref{boolsystem}) to avoid computation of the single equivalent equation $F=0$ as well as computation of the resolution product. We show in this paper that the ON expansion based algorithm can be constructed for satisfiability with these advantageous features for systems of Boolean equations whose functions belong to $B_{0}(X)$. Apart from these benefits we show that the algorithm has advantages such as
\bnum
\item The algorithm is applicable for decomposition of systems without any special representation of the problem such as in CNF or DNF. For CNF satisfiability the algorithm can provide a decomposition which generalizes the splitting rule of the well known DPLL algorithm. 
\item The algorithm in principle computes all solutions of the system whenever they exist otherwise it returns un-satisfiability.
\item The algorithm provides a natural decomposition of a bigger system into smaller systems to be solved independently. Hence this algorithm has an inherent parallelism. 
\enum

\subsection{SAT literature, elimination and relations with known approaches}
Well known problems of CNF or DNF satisfiability, commonly referred as SAT problems are special cases of the Boolean satisfiability problem. SAT problems have been investigated in Computer Science for several decades due to their vast applications to Propositional Logic, Artificial Intelligence, software and hardware verification, Operational Research and in recent times to cryptology. The vast bibliographies of \cite{hand,crah} and topics covered show the breadth and depth of research carried out in theory and algorithms for Boolean satisfiability in particular for systems represented in CNF and DNF. Much of the progress in SAT theory and practice is based on the celebrated DPLL algorithm and more modern approaches such as GRASP \cite{hand} as well as complete and incomplete algorithms. Specialized methods for SAT are a focus of developments reported in \cite{crah,bard,mezm}. In cryptology and computer algebra, algorithms based on extension of Gaussian elimination called XL method and Grobner basis computation have attracted considerable attention \cite{bard} where Boolean systems are often represented in the polynomials in several variables with $B_{0}$ co-efficients considered as the binary field $GF(2)$. Consistency of Boolean systems of equations discussed in \cite{rude,brow} on the other hand explain the method of elimination of variables. Elimination of variables in CNF SAT problems is equivalent to resolution of clauses. Due to high memory requirements early algorithms for CNF SAT which used resolution were modified to avoid resolution leading to the modern version known as DPLL. GRASP on the other hand is a different approach than DPLL which explores assignment substitutions along with incorporation of conflict resolving clauses.

ON expansion based algorithm developed in this paper is essentially equivalent to assignment substitutions decided by an ON set of functions. In certain SAT algorithms assignments of unknown variables are used to reduce the problem size, but can create conflicts which are resolved by going back and correcting the assignments. In ON expansion based assignments no conflicts arise. These are also equivalent to a several variable analogue of the splitting rule of DPLL at a single variable. One of the areas of current research in SAT algorithms and applications is development of parallel algorithms which are scalable for large data problems and for large number of processors. Recent survey \cite{hamw} shows the state of the art and challenges in development of parallel methods for SAT. An important direction in parallel algorithm development is decomposition of the Boolean system (or data set of CNFs or DNFs in SAT problems) for parallel processing. An advantages of ON based algorithm, as will be clear from this paper, is the natural decomposition of the problem it provides due to the ON expansion. Another issue with SAT algorithms is that they essentially compute only one solution if such exists or return un-SAT result. ON based algorithm on the other hand can compute all solutions when satisfiability holds, characterized by supports of the ON functions used. As an illustration we show computation of all rational solutions of an elliptic curve in a finite field.

\section{Zero sets of Boolean functions and correspondence with algebra}
Before discussing the consistency conditions we shall briefly present background results which shall be useful later at several places. The set $B_{0}(X)$ of functions in $n$ variables $X$ is a Boolean algebra w.r.t. Boolean operations in functions: For $f,g$ in $B_{0}(X)$ Boolean operations $\{+,.,'\}$ are defined with the help of values of functions in the Boolean algebra $B_{0}$ (whose Boolean operations are denoted by same symbols) on points $x$ in $B_{0}^n$ as follows:
\[
\begin{array}{rcl}
(f+g)(x) & = & f(x)+g(x)\\
(fg)(x) & = & f(x)g(x)\\
f'(x) & = & f(x)'\\
(f\oplus g)(x) & = & f(x)\oplus g(x)
\end{array} 
\]
For the Boolean inequality $f\leq g$ in $B_{0}(X)$ there is the following equivalence.
\[
\begin{array}{rrcll}
 & f & \leq & g &\\ \Leftrightarrow & f(x)g(x)' & = & 0 & \forall x\in B_{0}^{n}
\end{array}
\]
We now relate the Boolean functions with their zeros and supports which are formally denoted as follows:
\begin{enumerate}
\item The \emph{zero set} of a function $f$ in $B_{0}(X)$, denoted $V(f)$, is the set,
\[
V(f)=\{x\in B_{0}^{n}|f(x)=0\}
\]
\item The \emph{support} of a function $f$, denoted $\supp f$ is the set
\[
\supp f=\{x\in B_{0}^{n}|f(x)=1\}
\]
$\supp f$ is also often called the \emph{set of one values} in the literature \cite{crah}.
 \end{enumerate}
For $f$ in $B_{0}(X)$ we have the obvious identity with respect to set complement in $B_{0}^{n}$
\begin{equation}\label{vceqsupp}
V(f)^{c}=\supp f
\end{equation} 

The relationship of zeros with the algebraic properties is given by the following

\begin{prop}\emph{ For $f,g$ in $B_{0}(X)$
\beq \label{Valgebrarel}
\begin{array}{rrcl}
1) & V(f+g) & = & V(f)\cap V(g)\\
2) & V(f') & = & V(f)^{c}=\supp f\\
3) & V(fg) & = & V(f)\cup V(g)\\
4) & f\leq g & \mbox{iff} & V(g)\subset V(f)\\
& &  \Leftrightarrow & \supp f\subset\supp g
\end{array}
\eeq
}
\end{prop}

\bpf\
1) $V(f+g)=\{x\in B_{0}^{n}|f(x)+g(x)=0\}=\{x\in B_{0}^{n}|f(x)=g(x)=0\}=V(f)\cap V(g)$. 

2) $V(f')=\{x\in B_{0}^{n}|f(x)'=0\}=\{x\in B_{0}^{n}|f(x)=1\}=\supp f=V(f)^{c}$.

3) $V(fg)=V((f'+g')')=V(f'+g')^{c}=(V(f')\cap V(g'))^{c}=V(f')^{c}\cup V(g')^{c}=V(f)\cup V(g)$.

4) This is established in following equivalences
\[
\begin{array}{rcl}
f\leq g & \Leftrightarrow & fg'=0\mbox{ in $B_{0}(X)$}\\
 & \Leftrightarrow & f(x)g(x)'=0\forall x\in B_{0}^{n}\\
 & \Leftrightarrow & f(x)\leq g(x)\forall x\in B_{0}^{n}\\
 & \Leftrightarrow & \supp f\subset \supp g\\
 & \Leftrightarrow & V(g)=(\supp g)^{c}\subset (\supp f)^{c}=V(f)
\end{array}
\]
\epf

A point $a=(a_{1},\ldots,a_{n})$ in $B_{0}^n$ is the zero set $V(f_{a})$ of the function
\beq \label{Vfa}
\begin{array}{rcl}
f_{a}(x_{1},\ldots, x_{n}) & = & \sum_{i=1}^{n}(x_{i}\oplus a_{i})\\
 & = & \sum_{i=1}^{n}(x_{i}a_{i}'+x_{i}'a_{i})\\
 & = & \prod_{i=1}^{n}(x_{i}+a_{i})(x_{i}'+a_{i}')\\
\end{array}
\eeq
For points in $B_{0}^n$ we have another involutionary operation \emph{star} defined as follows: Let $a=(a_{1},\ldots,a_{n})$ then $a^{*}=(a_{1}',\ldots,a_{n}')$. Let $b=(b_{1},\ldots,b_{n})$ be another point in $B_{0}^n$. The \emph{dual} of a function $f(X)$ denoted $f^{d}(X)$ is defined as
\beq \label{dualfn}
f^{d}(X)=f(X^{*})'
\eeq
We have following relationships which can be proved easily.
\beq \label{fnstarrel}
\begin{array}{rrcl}
1) & a\cup b & = & V(f_{a}f_{b})\\
2) & a^{*} & = & V(((f_{a})^{d})')=\supp (f_{a})^{d}\\
3) & V(f)^{*} & = & \supp f^{d}
\end{array}
\eeq

Finally, for convenience we recall some of the well known properties of co-efficients in ON expansions. For Boolean functions $f,g$ in $B(X)$, an ON set of functions $\Phi$ and ON expansions (\ref{ONexp}) for $f$ and
\[
g(X)=\sum_{\phi_{i}\in\Phi}\beta_{i}(X)\phi_{i}(X) 
\]
following identities hold \cite{brow}
\begin{eqnarray}
f(X)+g(X) & = & \sum_{\phi_{i}\in\Phi}(\alpha_{i}(X)+\beta_{i}(X))\phi_{i}(X)\\
f(X)g(X) & = & \sum_{\phi_{i}\in\Phi}(\alpha_{i}(X)\beta_{i}(X))\phi_{i}(X)\\
f(X)' & = & \sum_{\phi_{i}\in\Phi}\alpha_{i}(X)'\phi_{i}(X)\\
f(X)\oplus g(X) & = & \sum_{\phi_{i}\in\Phi}(\alpha_{i}(X)\oplus\beta_{i}(X))\phi_{i}(X)
\end{eqnarray}

Composition of Boolean functions can also be expressed in ON expansion as follows. The simple proof is omitted. Let $g_{i}(X)$ for $i=1,\ldots,n$, $f(X)$ be Boolean functions in $n$-variables and
\[
g_{i}(X)=\sum_{j}\beta_{ij}(X)\phi_{j}(X)
\]
be expansions of $g_{i}(X)$ in the ON set as above. Then
\[
f(g_{1}(X),\ldots,g_{n}(X))=\sum_{j}\alpha_{j}(\beta_{1j}(X),\ldots,\beta_{nj}(X))\phi_{j}(X)
\]
This identity can be arrived at from previous identities for elementary Boolean operations between functions.

\section{Extension of the consistency condition}
We now take up the problem of extending the result of consistency of $f(X)=0$ obtained in \cite{sule} to the general case when the function $f$ need not be in $B(\Phi)$ for an ON set in $B(X)$. But we take up the special case in which $B$ is itself of the form $B_{0}(X_{2})$ hence the satisfiability is in terms of $\{0,1\}$ assignments. Although mathematically this is still a greatly restrictive formulation, it is nevertheless of considerable interest to Computer Science. In order to highlight the problem of consistency, consider a function $f(X)$ and $\Phi$ an ON set of functions of order $m$ both in $B_{0}(X)$. Consider an ON expansion (\ref{ONexp}) of $f(X)$. The questions that we want to investigate in this section are, 
\begin{quote}
\emph{
"What conditions on the co-effcieint functions $\alpha_{i}(X)$ arising in the expansion are necessary and/or sufficient for consistency of $f(X)=0$? How do such conditions change with the ON set? Further, since the expansion co-efficient functions $\alpha_{i}(X)$ are not unique whether such conditions can be more advantageous as compared to any selelction for a special choice of these co-efficients?"
}
\end{quote}
Note that these questions were answered for the special case when $f$ belonged to $B(\Phi)$ for a general Boolean algebra $B$ in \cite[Theorem 2, Corollary 1, Corollary 4]{sule} but in the present case we are considering general functions $f(X)$ in $B_{0}(X)$ hence the functions $\phi_{i}(X)$ in the ON set and the co-efficient functions $\alpha_{i}(X)$ may share common variables. We begin first with a general case for which we have an obvious result,

\begin{prop}\label{extensionofCon}\emph{Following two conditions hold in respect of an ON expansion (\ref{ONexp})
\begin{enumerate}
\item (Necessity). If $f(X)=0$ is consistent then there exists an index $i$ such that 
\beq\label{NecCon}
\alpha_{i}(X)=0
\eeq
is consistent. 
\item (Sufficiency). If the equation in co-efficients
\beq\label{SuffCon}
\sum_{i=1}^{m}\alpha_{i}(X)=0
\eeq
is consistent then $f(X)=0$ is consistent.
\end{enumerate}
}
\end{prop}

\bpf\
Sufficiency is obvious from the ON expansion for if there exists $x$ in $B_{0}^{n}$ such that $\alpha_{i}(x)=0$ for all $i$ then $f(x)=0$.

We thus prove the necessary condition. Let $f=0$ be consistent, then there exists $x$ in $B_{0}^{n}$ such that
\[
\sum_{i=1}^{m}\alpha_{i}(x)\phi_{i}(x)=0
\]
As $\Phi$ is ON,
\[
\sum_{i}\phi_{i}(X)=1
\]
hence there exists an index $i$ such that $\phi_{i}(x)=1$ and $\phi_{j}(x)=0$ for $j\neq i$. This implies
\[
0=f(x)=\alpha_{i}(x)
\]
and proves the necessary condition. 
\epf

The proposition implies that consistency of at least one equation $\alpha_{i}(X)=0$ for some $i$ is necessary for consistency of $f=0$ while consistency of the system
\[
\alpha_{i}(X)=0\forall i
\]
is sufficient. 

Consistency of an equation $f=0$ has important connection with the theory of \emph{elimination} of Boolean variables. (We refer the reader to \cite{rude, brow} for a background on elimination theory). In the analysis of elimination of variables in Boolean equations, the Boole-Shannon expansion in terms of a single variable $x$ is
\[
f(X)=f_{1}x+f_{0}x'
\]
(where $f_{1}=f(x=1)$, $f_{0}=f(x=0)$) which is an ON expansion relative to the ON set $\Phi=\{x,x'\}$. In this case the function $f$ belongs to the special class $B(\{x,x'\})$ where $B=B_{0}(X-\{x\})$. Hence the necessary and sufficient condition of \cite{sule} is the same as the well known consistency condition in terms of the eliminant
\[
f_{0}f_{1}=0
\]
In the present general case we do not have co-efficients $\alpha_{i}$ in the expansion independent of the variables $X$ in the function, nor do we have a necessary and sufficient condition for consistency in general. Hence the conditions of proposition \ref{extensionofCon} cannot be related to elimination of variables. But if the ON set is a set of minterms in a subset of variables in $X$ (or all of these variables), we can get the necessary and sufficient condition as shown in the next subsection and can be related to elimination. Hence above conditions for consistency appear to be mainly of theoretical consequence and do not provide any computationally useful consequence which was the main motivation of this paper. Also the proof explicitly makes use of the fact that $f$ belongs to $B_{0}(X)$. Analogous conditions can be derived for the one value form of the equation as follows.

\begin{cor}\emph{Following two conditions hold in respect of an ON expansion (\ref{ONexp})
\begin{enumerate}
\item (Necessity). If $f(X)=1$ is consistent then there exists an index $i$ such that 
\beq
\alpha_{i}(X)=1
\eeq
is consistent. 
\item (Sufficiency). If the equation in co-efficients
\beq
\prod_{i=1}^{m}\alpha_{i}(X)=1
\eeq
is consistent then $f(X)=1$ is consistent.
\end{enumerate}
}
\end{cor}

\bpf\
$f(X)=1$ is consistent iff $f(X)'=0$ is consistent. The ON expansion of $f(X)'$ in $\Phi$ is
\[
f(X)'=\sum_{i=1}^{m}\alpha_{i}(X)'\phi_{i}(X)
\]
Hence both conditions follow from the conditions (\ref{NecCon}), (\ref{SuffCon}) respectively.
\epf

\subsection{Special expansion: ON set of minterms}
Consider now the special case when the variables $X$ are partitioned in two subsets $X_{1}$, $X_{2}$ and the ON set $\Phi=\{\mu_{i}(X_{1})\}$ is a set of minterms $\mu_{i}$ in $X_{1}$ variables. For such an ON set an expansion of $f$ in $B_{0}(X)$ of the form
\[
f(X)=\sum_{i=1}^{m}(f/\mu_{i})(X_{2})\mu_{i}(X_{1})
\]
can be chosen in which the co-efficients of expansion $\alpha_{i}=(f/\mu_{i})$ are functions of the $X_{2}$ variables alone. For such a special expansion and ON functions we can prove a necessary and sufficient condition for consistency as follows.

\begin{theorem}
\emph{For a partition of $X$ as described above, the special ON set $\Phi$ of minterms in $X_{1}$ and the choice of a special expansion of $f$ as above, the equation $f(X)=0$ is consistent iff
\[
\prod_{i=1}^{m}\alpha_{i}(X_{2})=0
\]
is consistent.
}
\end{theorem} 

\bpf\
If $f(X)=0$ is consistent then by the necessary condition of proposition \ref{extensionofCon} there exists an index $i$ such that
\[
\alpha_{i}(X_{2})=0
\]
is consistent. This implies the above condition.

Conversely, if $f(X)=0$ is not consistent, then
\[
f(x)=1 \forall x\in B_{0}^{n}
\]
Let $X=(X_{1},X_{2})$ denotes the $X_{1}$ and $X_{2}$ variable components. For each assignment $x_{1}$ there is an $i$ and a unique minterm $\mu_{i}(X_{1})$ such that $\mu_{i}(x_{1})=1$ while $\mu_{j}(x_{1})=0$ for $j\neq i$. Hence
\[
f(x_{1},X_{2})=\alpha_{i}(X_{2})
\]
which implies that
\[
\alpha_{i}(x_{2})=1\forall\, x_{2}
\]
for all assignments $x_{2}$. As $x_{1}$ is varied over all assignments it follows that $\alpha_{i}(x_{2})=1$ for all $i$ and $x_{2}$. Hence
\[
\prod_{i=1}^{m}\alpha_{i}(x_{2})=1
\]
which implies that
\[
\prod_{i=1}^{m}\alpha_{i}(X_{2})=0
\]
is not consistent. This proves sufficiency.
\epf

\subsection{Computationally useful formulation of consistency}
Finally, we consider a formulation of the consistency condition in terms of an ON expansion which will be useful for the purpose of computation as well as resolving the problem of consistency and computation of solutions of systems of equations. 

\begin{prop}\label{CorComputational} \emph{Given an ON expansion (\ref{ONexp}) of $f(X)$ the equation $f(X)=0$ is consistent iff there exists an index $i$ in $1,\ldots,m$ such that the system
\[
\begin{array}{rcl}
\alpha_{i}(X) & = & 0\\
\phi_{i}(X) & = & 1
\end{array}
\]
is consistent.}
\end{prop}

\bpf\
If the equation is consistent and $x$ in $B_{0}^{n}$ is a solution then
\[
\sum_{i=1}^{m}\alpha_{i}(x)\phi_{i}(x)=0
\]
But as $\phi_{i}$ are ON this implies there exists $i$ in $[1,m]$ such that $\phi_{i}(x)=1$ and that $\phi_{j}(x)=0$ for $j\neq i$. This implies that the system above is consistent.

To prove sufficiency, let the above condition hold for some index $i$ and let $q$ in $B_{0}^n$ be a solution of the equations. Then, since $\phi_{j}(X)\phi_{i}(X)=0$ for all $j\neq i$ evaluating this identity at $q$ gives $\phi_{j}(q)=0$ for all $j\neq i$. Hence $f(q)=0$ since all terms in the (\ref{ONexp}) when evaluated at $q$ are zero. This proves that the condition is sufficient for consistency of the equation $f(X)=0$.
\epf

This formulation of consistency is important for two reasons. First, it gives a necessary and sufficient condition in the general case, second it relates consistency of the equation $f=0$ to the consistency of a system in which one equation is in a member of the ON set. Since ON functions are well characterized \cite{rude} their solutions and one values can also be assumed well characterized, hence the system indicates a natural decomposition of the space of search for solutions. This is essentially a computational advantage and shall be developed further to tackle the problem of solving systems of Boolean equations efficiently.

\subsection{Conjugate consistency problems}
So far we have considered the consistency of the equation $f(X)=0$ in terms of an ON expansion of $f$. There is however the associated function $f(X^{*})$ in the definition of the dual function $f^{d}$ whose support is the conjugate of the set of solutions $V(f)$ of this equation as illustrated in equation (\ref{fnstarrel}). Hence it is natural that instead of considering the problem of solving one equation $f(X)=0$ associated problems are also explored and a symmetry is achieved in some sense. We shall call these as \emph{conjugate problems}. The ON functions and expansion co-efficients of conjugate problems are also closely related as shown below.

Let $\Phi=\{\phi_{1},\ldots,\phi_{m}\}$ be an ON set of functions in $B_{0}(X)$ then the set 
\[
\Phi*=\{\phi*_{1},\ldots,\phi*_{m}\}=\{\phi_{1}(X^{*}),\ldots,\phi_{m}(X^{*})\}
\]
is also ON. To the ON expansion (\ref{ONexp}) there is an associated expansion
\begin{equation}\label{conjONexp}
f(X^{*})=\sum_{i=1}^{m}\alpha_{i}(X^{*})\phi*_{i}(X)
\end{equation}
We shall call this a \emph{conjugate expansion}. The supports of $\phi_{i}(X)$ and $\phi_{i}(X^{*})$ are also conjugates i.e. for a $q$ in $B_{0}^{n}$,
\[
\phi_{i}(q)=1\mbox{ iff }\phi*_{i}(q^{*})=1
\]

The Boolean equation $f(X)=0$ has the solution set $V(f)$ in $B_{0}^{n}$ while from $3)$ of (\ref{fnstarrel}) it follows that $V(f)^{*}$ is the zero set of $f(X^{*})$. We shall call such problems whose solution sets are conjugates as \emph{conjugate consistency problems}. For a single equation we have

\begin{prop}\label{CojugateProb1}\emph{Following pairs of consistency problems are conjugate with solution sets indicated
\[
\begin{array}{rrl}
(1) & V(f): f(X)=0 & V(f)^{*}: f(X^{*})=0\\
(2) & V(f)^{c}:f(X)=1 & V(f)^{*c}:f(X^{*})=1  
\end{array}
\]
 } 
\end{prop}
          
\section{Elimination of variables in systems case}
Let $f$ be a function of one variable $\xi$ in $B(\xi)$. The equation $f(\xi)=0$ is consistent iff $f(1)f(0)=0$ in $B$ \cite[Section 7.3.1]{brow}. This leads to the notion of elimination of variables from Boolean equations. Elimination of a variable $\xi$ in an equation $f(\xi,Y)=0$ where $Y$ are other variables in the Boolean function $f$ in $B(\xi,Y)$ follows from the expansion
\[
f(\xi,Y)=f(1,Y)\xi+f(0,Y)\xi'
\]
From the consistency condition it follows that $f(\xi,Y)=0$ is consistent iff
\[
f(1,Y)f(0,Y)=0
\]
is consistent in $B(Y)$ which is a new equation involving only $Y$ variables. The function $F(Y)=f(1,Y)f(0,Y)$ is called the \emph{eliminant} of $f$ w.r.t. $\xi$. 

Now if instead of $f=0$ there is a system of equations in variables $x,Y$ where $x$ is a single variable and other variables $Y$,
\begin{eqnarray}\label{systemfg}
f(x,Y) & = & 0\\g(Y) & = & 0
\end{eqnarray}
then we have the consistency condition given by

\begin{lemma}{\em
The system (\ref{systemfg}) is consistent iff $F(Y)+g(Y)=0$ is consistent where $F(Y)$ is the eliminant of $f(x,Y)$ w.r.t. $x$.
}
\end{lemma}

\bpf\
The system (\ref{systemfg}) is consistent iff the single equation
\[
f(x,Y)+g(Y)=0
\]
is consistent. By elimination of $x$, this equation is consistent iff
\[
\begin{array}{rcl}
[f(1,Y)+g(Y)][f(0,Y)+g(Y)] & = & f(1,Y)f(0,Y)+(f(1,Y)+f(0,Y))g(Y)+g(Y)\\
 & = & F(Y)+g(Y)=0
\end{array}
\]
is consistent (which is equivalent to consistency of the system $F(Y)=0,g(Y)=0$) where $F$ is the eliminant of $f(x,Y)$ w.r.t. $x$. 
\epf

Thus for the system (\ref{systemfg}) where the $x$ variable is absent in $g$, elimination of $x$ can be carried out to get the equation $F(Y)=0$ independent of computation on $g$. However the consistency condition has another interpretation

\begin{cor}\emph{The system (\ref{systemfg}) is consistent iff $g(Y)=0$ is consistent and there exists a solution $\alpha$ of $g(Y)=0$ such that the equation $f(x,\alpha)=0$ is consistent.} 
\end{cor}

\bpf\
Since $F(Y)=f(1,Y)f(0,Y)$ consistency of $F(Y)+g(Y)=0$ is equivalent to $g(Y)=0$ being consistent and there exists $\alpha$ such that $g(\alpha)=0$ and $f(1,\alpha)f(0,\alpha)=0$ which is the consistency of $f(x,\alpha)=0$.
\epf

Hence alternatively consistency of the system (\ref{systemfg}) can be obtained by evaluating $f$ at solutions of $g$ and determining consistency of the resultant function. This has computational implications which are discussed next.

\subsection{Computation of solutions without converting to single equation form}
Above corollary has an advantageous implication for computation of consistency and a solution of systems. For a system of equations above in which the variables are partitioned as in (\ref{systemfg}) in $X,Y$ where $|X|$ is large, if the number of variables $Y$ is small enough or the consistency and computation of solutions of $g(Y)=0$ is easy, then the consistency of the system can be determined by decomposing the system into independent problems of consistency of $f(X,\alpha)=0$ and $g(\alpha)=0$. Having the variables $Y$ assigned values $\alpha$ computing consistency in $X$ variables is considerably advantageous than elimination of $X$ with indeterminate $Y$. Such an approach is thus an alternative to elimination of variables $X$ in deciding consistency and finding a solution without converting the problem to a single equation. We shall show that ON decomposition of functions in a general system of equations exploits this aspect in computation and is hence expected to provide a decomposition of systems without converting them to single equation form. Following proposition shall be a preliminary result before building up our algorithm for general systems.

We now consider the special case of functions $B_{0}(X)$ in $n$-variables. Let $\Phi$ be a set of $m$ ON functions and consider an expansion (\ref{ONexp}) of $f$. For any non zero function $\phi$ in $\Phi$, there exists a non-empty set $Q$ in $B_{0}^n$ such that
\[
\phi(q)=1\forall q\in Q
\]
the union of all such sets is called the set of one values or \emph{support} of $\phi$ denoted $\supp\phi$. Following proposition is almost a restatement of proposition \ref{CorComputational}.

\begin{prop}\label{coninalpha}\emph{ 
Let $f$ be a Boolean function given with an ON expansion (\ref{ONexp}). Then $f(X)=0$ is consistent iff there exists an index $i$ and an assignment $q$ in support of $\phi_{i}$ such that $\alpha_{i}(q)=0$. Every solution of $f(X)=0$ when consistent arises this way.}
\end{prop}

\bpf\
If $f(X)=0$ is consistent there exists a solution $q$ in $B_{0}^n$ which satisfies
\[
\alpha_{i}(q)\phi_{i}(q)=0\forall\, i=1,\ldots,m
\]
Since $\sum\phi_{i}=1$ there is an index $i$ such that $\phi_{i}(q)=1$ hence $\alpha_{i}(q)=0$. This proves necessity and also shows that every solution $q$ satisfies these conditions.

Conversely, if such an index $i$ and $q$ satisfying given conditions exist such that $\phi_{i}(q)=1$ then $\phi_{j}(q)=0$ for all $j\neq i$. Hence since $\alpha_{i}(q)=0$
\[
f(q)=\alpha_{i}(q)+\sum_{j\neq i}\alpha_{j}(q)\phi_{j}(q)=0
\]
Hence $f=0$ is consistent and $q$ is a solution. This shows the conditions are necessary for consistency and every such $q$ is a solution.
\epf

Computational consequence of this proposition is quite direct. Since $\alpha_{i}$ satisfy
\[
\alpha_{i}(X)\phi_{i}(X)=f(X)\phi_{i}(X)
\]
it follows that
\[
\alpha_{i}(q)=f(q)
\]
for $q$ in $\supp\,\phi_{i}$. Hence to determine consistency of $f=0$ we can search over index $i$ and the support of $\phi_{i}$, independently for each $i$, such that $f(q)=0$.  We write this explicitly as a corollary to the above proposition.

\begin{cor}\label{solnviacoef}\emph{$f(X)=0$ is consistent iff there exists an index $i$ and an assignment $q$ in $\supp\phi_{i}$ such that $f(q)=0$.}
\end{cor}

This result observes that $f=0$ is consistent iff there is a zero of $f$ in the support of at least one $\phi_{i}$ in the ON set $\Phi$. This has important computational consequences depending on the choice of the ON set for computing solutions of a Boolean system (\ref{boolsystem}).

\subsection{Consistency of the Boolean system}
To write the consistency condition for the Boolean system (\ref{boolsystem}) we first express the individual functions $f_{i},g_{i}$ in ON expansion in terms of an ON set $\Phi$. Let these expansions be
\begin{equation}\label{ONexpofsyst}
\begin{array}{rcl}
f_{i}(X) & = & \sum_{j=1}^{m}\alpha_{ij}(X)\phi_{j}(X)\\
g_{i}(X) & = & \sum_{j=1}^{m}\beta_{ij}(X)\phi_{j}(X)
\end{array}
\end{equation}
for $i=1,\ldots,N$.

For the case of systems following theorem shows that a search for solutions can be carried out without computing the single function $F$.

\begin{theorem}\label{Th:Solnassign}
\emph{Let a Boolean system (\ref{boolsystem}) has ON expansions (\ref{ONexpofsyst}) of its functions.Then the system (\ref{boolsystem}) is consistent iff there exists an index $k$ in $[1,m]$ and an assignment $q$ in $\supp\phi_{k}$ such that
\[
\alpha_{ik}(q)=\beta_{ik}(q)\forall i=1,\dots,N
\]
Every solution of a consistent system (\ref{boolsystem}) arises this way.}
\end{theorem} 

\bpf\
Consider the single equation $F(X)=0$ equivalent to the system (\ref{boolsystem}) and the ON expansions of the individual functions (\ref{ONexpofsyst}) where 
\[
\begin{array}{rcl}
F(X) & = & \sum_{i=1}^{N}(f_{i}(X)\oplus g_{i}(X))\\
 & = & \sum_{j=1}^{m}\sum_{i=1}^{N}(\alpha_{ij}(X)\oplus \beta_{ij}(X))\phi_{j}(X)
\end{array}
\]
Using the identities above for ON expansion w.r.t. $\Phi$ and  the result of proposition \ref{coninalpha} it follows that when $F(X)=0$ is consistent and $q$ is a solution, there is an index $k$ in $[1,m]$ such that $q$ belongs to $\supp\phi_{k}$ and satisfies
\[
\sum_{i=1}^{N}(\alpha_{ik}(q)\oplus\beta_{ik}(q)=0
\]
which proves necessity of the condition and shows that evevry solution of the system satisfies this condition.

Conversely let there exists $k$ in $[1,m]$ and a $q$ in $\supp\,\phi_{k}$ which satisfies
\[
\alpha_{ik}(q)=\beta_{ik}(q)\forall\, i=1,\ldots,N
\]
then from the (\ref{ONexpofsyst}) and noting that $\phi_{i}(q)=0$ for $i\neq k$ it follows that
\[
f_{i}(q)=g_{i}(q)\forall\,i=1,\ldots,N
\]
Hence $q$ is a solution of the system (\ref{boolsystem}) which proves sufficiency and shows that every solution of a consistent system arises this way.
\epf

\begin{remark}\emph{The proof above shows that once the ON set $\Phi$ is chosen the search space of solutions can be decomposed to subsets $\supp\,\phi_{i}$ and further the resultant assignments satisfying the system can be found by evaluation and checking whether $f_{i}(q)=g_{i}(q)$. Hence the computation of co-efficient functions $\alpha_{i}$, $\beta_{i}$ is not required.}
\end{remark}

This theorem with the above remark forms a basis of a computational procedure developed in the next which is useful for writing parallel computational algorithms.

\subsection{Computational procedure for Boolean systems}
Solving Boolean systems by a scalable process is the central goal of practical computation. By scalable it is meant that the process works efficiently even when the size of the system is large enough to solve real industrial problems, as well as is able to utilize multiple parallel computations and works efficiently even over large number of such computing nodes (as are available in current technology). ON expansion based consistency condition developed above helps achieve scalability since the consistency of the original system and its solutions are determined independently from decomposed systems after substituting the assignments from supports of the ON functions. Before writing this process formally as an algorithm we need to take into account assignments arising from or solutions of some of the simplest Boolean systems which need not be expanded by ON sets to solve them and the system can be reduced after such trivial assignments. As a partial list of such systems and reductions consider
\begin{enumerate}
\item Number of variables $n=|X|$ as well as equations $N$ in (\ref{boolsystem}) is small. In such a case all solutions and consistency can be searched over $B_{0}^{n}$ by brute force search.
\item Equations of the type $l_{i}=l_{j}$ in literals. If such an equation arises one variable in the system is reduced.
\item Unit clauses such as $l=1$, trivial equations such as,
\[
\sum l_{i}=0, \prod l_{i}=1
\]
all are equations for which assignments are trivially determined.
\end{enumerate}
We shall denote the function $\mbox{Trivsolve}()$ as a generic function which makes trivial assignments of variables whenever possible thereby reducing the system to a new system in which assigned variables are removed. This reduced system is considered an output of this function. The algorithm which carries out ON decomposition can now be written as follows.

\begin{algorithm}[Decomposition]\emph{$\mbox{Decompose()}$
\begin{enumerate}
\item \textbf{Input} System denoted by $(S,X)$ as in (\ref{boolsystem}) with variables $X$. $n_{0}$ the largest number of variables below which the solution of $S$ or its inconsistency can be determined by brute force search.
\item $n=|X|$, \textbf{while} $n>n_{0}$ \textbf{repeat}
\item Choose an ON set of $\Phi=\{\phi_{i},i=1,\ldots,m\}$
\item For each $i$ and $q$ in $\mbox{supp}\,\phi_{i}$ determine the collection of all systems $S(q)$. Determine the variables $X(q)$ of $S(q)$.
\item Distribute each of these systems to an independent node for independent computation
\[
(S,X)\leftarrow (S(q),X(q))
\]
\item \textbf{return} system to be solved: $(S,X)$.
\end{enumerate} 
}
\end{algorithm}

The main algorithm is now as follows

\begin{algorithm}[ON decomposition based solver]\emph{$\mbox{BoolSolve}()$
\begin{enumerate}
\item Input $(S,X)$, $n_{0}$
\item $n=|X|$ \textbf{while} $n>n_{0}$ \textbf{repeat}
\item $\mbox{TrivSolve}(S,X)$
\item Decompose$(S,X)$
\item At each node \textbf{if} $n\leq n_{0}$ \textbf{return} solution of the system or set flag $SAT=F$, broadcast flag.
\end{enumerate}
}
\end{algorithm}

The ON expansion thus plays the role of decomposing the original system to a smaller size for independent parallel computation. In practice much of the efficiency can be gained by heuristics in deciding the ON set as well as in strategies for reducing variables during trivial solutions.

\section{Decomposition and reduction of systems at partial assignments}
In a general Boolean algebra $B$, a characterization of ON sets is obtained in \cite[Theorem 4.2]{rude} which is reproduced below for convenience. If $\{y_{1},y_{2},\ldots,y_{m}\}$ is an ON set in $B$ then there are elements of the form $u_{1},\ldots,u_{m-1}$ in $B$ such that
\[
\begin{array}{rcll}
y_{1} & = & u_{1}' &\\
y_{j} & = & u_{1}u_{2}\ldots u_{j}' \mbox{ for }2\leq j\leq m-1\\
y_{m} & = & u_{1}u_{2}\ldots u_{m-1}
\end{array}
\]
From this characterization of ON sets it can be shown as in \cite[theorem 4.2, corollary]{rude} that ON sets $\Phi$ of any order $m$ from $2$ to $2^n$ exist and are related to the partition of the set $\{0,2^{n-1}\}$. Let symbols $\mu$ denote minterms in $n$ variables which are of the form
\[
\mu=x_{1}^{a_1}\ldots x_{n}^{a_n}
\]
where $a_{i}=0,1$ and the literals satisfy $x^{1}=x$, $x^{0}=x'$ there are $2^{n}$ minterms which can be indexed by the set $\{0,1,\ldots,2^{n-1}\}$. Consider 
\[
\cup_{j=1}^{m} M_{j}
\] 
to be a partition of the set $\{0,1,\ldots,2^{n-1}\}$. Then the set $\Phi=\{\phi_{i},i=1,\ldots,m\}$ defined by
\[
\phi_{i}(X)=\sum_{j}\mu_{ij},j\in M_{i}
\]
where $\mu_{ij}$ are minterms indexed according the the partition above is ON. Conversely to any ON set $\Phi$ there is a unique partition as above of $\{0,1,\ldots,2^{n-1}\}$ which defines the functions $\phi_{i}$ in $\Phi$. To explain these representations of ON sets we consider illustrative examples.

\begin{example}\emph{In three variables $x,y,z$ consider ON sets
\begin{enumerate}
\item Of order $2$: $x,x'$, represented in minterms as
\[ 
\begin{array}{rcl}
x & = & xyz+xy'z+xyz'+xy'z'\\x' & = & x'yz+x'y'z+x'yz'+x'y'z'
\end{array}
\]
\item Of order $3$: $x,x'y,x'y'$, represented in minterms as
\[
\begin{array}{rcl}
x & = & xyz+xy'z+xyz'+xy'z'\\x'y & = & x'yz+x'yz'
\end{array}
\]
\item Of order $4$: $x,x'y,x'y'z,x'y'z'$,
\item As there are $8$ minterms other ON sets can be constructed by summing the minterms.
\end{enumerate}
}
\end{example}

Due to this characterization of ON sets as sums of minterms the solution sets of equations of the form
\[
\phi_{i}(X)=1
\]
is trivial and can be described as follows.

\begin{prop}\emph{Let $\Phi$ be an ON set of order $m$ in $B(X)$ and $\phi_{i}$ an element, the equation $\phi_{i}(X)=1$ is consistent. A string $q$ in $B^n$ satisfies $\phi_{i}(q)=1$ iff there exists a minterm $\mu_{ij}\in M_{i}$ such that $\mu_{ij}(q)=1$. All solutions of $\phi_{i}(q)=1$ arise this way.}
\end{prop}

\bpf\
Consider first proving the second and third statements. Note that the ON set being reduced has no zero functions among its elements. Since
\[
\phi_{i}=\sum_{j\in M_{i}}\mu_{ij}
\]
for minterms $\mu_{ij}$, if there is $q$ such that $\phi_{i}(q)=1$, then this implies $\mu_{ij}(q)=1$ for some $j$. If $\mu_{ij}$ is the product  
\[
\mu_{ij}=x_{1}^{a_{1}}x_{2}^{a_{2}}\ldots x_{n}^{a_{n}}
\]
then $\mu_{ij}(q)=1$ has the only solution given by assignments $x_{j}=1$ if $a_{j}=1$ and $x_{j}=0$ otherwise. This proves the last two statements. Hence $\phi_{i}(q)=1$ is always consistent which proves first statement. 
\epf

ON sets of functions can be more generally created by products and sums of functions in partitions of ON sets as follows.

\begin{prop}\emph{Let $B$ be any Boolean algebra and $X=X_{1}\cup X_{2}$ where subsets $X_{1}$, $X_{2}$ may not be disjoint.
\begin{enumerate}
\item If $\Phi$ is an ON set in $B(X)$ of order $m$ and 
\[
\Phi=\cup_{i=1}^{t} F_{i}
\]
is a disjoint partition of $\Phi$. Then the set $\{\psi_{i}\}$ defined by
\[
\psi_{i}=\sum_{\phi_{i}\in F_{i}}\phi_{i}
\]
is an ON set of order $t$.
\item If $\Phi_{i}$ is an ON set in $B(X_{i})$ of order $m_{i}$ for $i=1,2$ then $\{\phi_{1k}\phi_{2l}\}$ for $\phi_{ik}$ in $\Phi_{i}$ is an ON set of order $m_{1}m_{2}$.
\end{enumerate}
}
\end{prop}

\subsection{ON terms and Partial assignments}
In general if $\Phi$ is an ON set of functions and $\phi$ is an element, as shown in above proposition there are multiple solutions to $\phi(q)=1$. However if the set $\Phi$ consists of ON terms $\{t_{1},t_{2},\ldots,t_{m}\}$ then these multiple solutions are characterized by unique partial assignments of variables defining $t_{i}$. For instance if $x',xy',xyz', xyz$ is an ON set then $xy'=1$ has all solutions given by $(x,y,z)=(1,0,0),(1,0,1)$. Thus the unique assignments of $x,y$ are enough to construct all solutions by assigning free variables ($z$ in this case) freely. For a term
\[
t(X)=\prod_{i}x_{i}^{r_{i}}
\]
The partial assignment defined by $t(X)=1$ is  $x_{i}=1$ when $r_{i}=1$ and $x_{i}=0$ when $r_{i}=0$. The well known concept of \emph{ratio} $f/t$ of a function $f(X)$ and a term $t(X)$ is defined as the function \cite{brow},
\[
(f/t)(X)=f(t(X)=1)
\]
where the partial assignments of variables in $t(X)$ are substituted in $f$ while $f/t$ is a function of the free variables which do not appear in the term $t$. As a useful notation for assignments of variables defined by partial assignments, consider a term $t(X)$ involving a subset of variables $X_{1}$ in $X$ and $X_{2}$ denote rest of the (free) variables. Denote the partial assignments (that of $X_{1}$) which satisfy $t(X)=1$ by $q(t)$. If a tuple $q_{2}$ are assignments of the variables $X_{2}$, denote by $(q(t)\Vert q_{2})$ the assignment of $X$ which is one of the solutions of $t(X)=1$. The assignments of individual variables $x_{i}$ at index $i$ are suppressed in this notation as this an algorithmic task which will be incorporated in the computation.  

Using partial assignments we can restate the consistency of systems (\ref{boolsystem}) from theorem \ref{Th:Solnassign} as follows.

\begin{cor}\label{Cor:ONtermexp}\emph{Let $T=\{t_{1},t_{2},\dots,t_{m}\}$ be an ON set of terms. The system (\ref{boolsystem}) is consistent iff there is an index $k$ in $[1,m]$ such that the system
\[
(f_{i}/t_{k})=(g_{i}/t_{k}),i=1,\ldots,N
\]
is consistent. If $q$ is a solution of this system, then $(q(t_{k})\Vert q)$ gives a solution of the system (\ref{boolsystem}). every solution of a consistent system (\ref{boolsystem}) arises in this form. 
}
\end{cor} 

\bpf\
The ON set is now $T$ of terms. Hence by the theorem there exists an index $k$ such that for $q$ in $\supp\,t_{k}(X)$ the equations are satisfied
\[
f_{i}(q)=g_{i}(q)\forall i=1,\ldots,N
\]
But $q$ arises as $(q(t_{k})\Vert q_{2})$ where $q_{2}$ satisfies
\[
(f_{i}/t_{k})(q_{2})=(g_{i}/t_{k})(q_{2})\forall i=1,\ldots,N
\]
This proves that the $(f_{i}/t_{k})=(g_{i}/t_{k})$ is consistent with solution $q_{2}$ and every solution is of this form.
\epf

As a compact notation let the system (\ref{boolsystem}) be denoted $S$ and the component systems in ON expansion in terms of $T$ be denoted $S/t_{k}$.

\subsection{An example of computing rational solutions}
We show an application of the ON term based procedure stemming from corollary \ref{Cor:ONtermexp} to computation of the roots of an algebraic equation over a finite field. Consider an elliptic curve expressed in Weierstrass form with co-efficients defined over the finite field $K=\ff_{2^3}$ by
\[
\begin{array}{rcl}
E & y^{2}+xy+x^{3}+(1+\theta)x^{2}+\theta=0
\end{array}
\]
where $K$ is described in a polynomial basis defined by the irreducible polynomial $\theta^{3}+\theta+1$. Hence we have $\theta^{3}=\theta+1$. We want to compute all solutions of this equations (points $(x,y)$ in $K^{2}$ constituing the elliptic curve $E$) in $K$. Note that it is not guranteed that $E$ is non empty.

A Boolean equation approach to solve this problem is to express variables $x$, $y$ in $K$ in  co-ordinates $x=a+b\theta+c\theta^{2}$, $y=d+e\theta+f\theta^{2}$ in $\ff_{2}$ treated as the Boolean algebra $B_{0}$. This defines a Boolean system of equations in unknowns $a,..., f$ in $B_{0}$. The solution set of this system thus gives all points on $E$ and when the equations are inconsistent $E$ is empty. Consider the ON set of terms
\[
T=\{c,c'b,c'b'a,c'b'a'\}
\]
When partial assignments defined by these terms are substituted in the equation of $E$ we get independent quadratic equations over $K$ such as
\[
pT^{2}+qT+r=0
\]
with $p\neq 0$. This equation has well known solutions. For $q=0$ the multiple solution $\sqrt{r/p}$. For $q\neq 0$, defining $s=pr/q^2$ solution exists iff $\mbox{Tr}\,s=0$. The two solutions are $u,u+1$ where $u$ is a nonzero solution in the kernel of the Artin-Schreier map $x\rightarrow x^{2}+x$. Computation of this kernel can be yet another Boolean system problem in the chosen basis for $K$. In this example we shall only show how different substitutions of partial assignments result into checking existence via trace evaluation as above.

Let the ON terms in $T$ be indexed as $t_{i},i=1,\ldots,4$. We denote substitutions by $E/t_{i}$
\begin{enumerate}
\item $E/t_{4}$. In this $x=0$. The solutions exists, $y=\sqrt{\theta}$.
\item $E/t_{3}$. In this $x=1$. The equation is $y^{2}+y+(1+\theta+\theta^{2})$. $\mbox{Tr}\,(1+\theta+\theta^{2})=1$, hence no solution.
\item $E/t_{2}$. In this $x=a+\theta$. Hence two cases are studied $x=\theta$ and $x=1+\theta$. In both cases the $s$ calculated has trace zero. Hence solutions exist for both $a=0,1$.
\item $E/t_{1}$. Here $x=a+b\theta+\theta^{2}$. Hence we can further expand the equation $E/t_{1}=0$ w.r.t. ON terms $a',ab',ab$. Let these be denoted by indexing $E/t_{11},E/t_{12},E/t_{13}$ respectively with values of $x$ as in the last column of the following table
\[
\begin{array}{rcl}
E/t_{11} & b=1 & (\theta+\theta^{2})\\ 
 & b=0 & \theta^{2}\\
E/t_{12} & & (1+\theta^{2})\\
E/t_{13} & & (1+\theta+\theta^{2})
\end{array}
\]
\end{enumerate}
Evaluation of traces of $s$ defined by $x$ in each of the substitutions of $x$ above gives all remaining solutions. 

\subsection{Algorithm for CNF-SAT}
Finally we shall discuss the ON expansion based algorithm for a CNF satisfiability problem. Let $C$ or $C(X)$ denote the set of all clauses over the set of literals denoted $X$ and $\mathcal{S}(C)$ denote the set of all assignments for satisfying all clauses in $C$. If $x$ in $X$ is a pure literal (i.e. no clause in $C$ contains $x'$) then $\mathcal{S}(C)$ is non empty iff for the set $C/x$ which equals $\{C_{j}(x=1)|C_{j}\in C\}$ the set $\mathcal{S}(C/x)$ is non empty. Let $t_{r+1}=x_{1}\ldots x_{r}$ be the term denoting product of all pure literals in $C$ and consider the ON set
\[
T=\{t_{1},\ldots,t_{r+1}\}
\]
where
\begin{equation}\label{purelitONterms}
\begin{array}{rcl}
t_{1} & = & x_{1}'\\
t_{2} & = & x_{1}x_{2}'\\
\vdots & = & \vdots\\
t_{r} & = & x_{1}x_{2}\ldots x_{r}'\\
t_{r+1} & = & x_{1}x_{2}\ldots x_{r}
\end{array}
\end{equation}
Then the following proposition follows.

\begin{prop}\emph{With the $r$ positive variables and ON terms defined above $C$ is satisfiable iff $C/t_{r+1}$ is satisfiable. All satisfying assignments of $C$ when it is satisfiable are obtained as the union of all assignments $\mathcal{S}(C/t_{i})$ along with the partial assignments $q(t_{i})$.}
\end{prop}

We shall illustrate this process by an example.

\subsubsection{Example of CNF-SAT 1}
Consider the CNF set $C$ given by the matrix whose first row gives column indices indexing the variables. Subseuqnt rows indicate clauses. In this notation $1$ below a column $i$ indicates $x_{i}$ and $-1$ indicates $x_{i}'$ term in the clause.
\[
\barr{rrrrrrrr}
1 & 2 & 3 & 4 & 5 & 6 & 7 & 8 \\
1 & 0 & -1 & 0 & 0 & 1 & 0 & 0\\
0 & 1 & -1 & 0 & 1 & 1 & 1 & 0\\
1 & 1 & -1 & 0 & -1 & -1 & 0 & 1\\
0 & -1 & 0 & 1 & 0 & 0 & -1 & -1\\
0 & 0 & 0 & -1 & 0 & 0 & 0 & 1
\earr
\]
Thus $x_{1},x_{3}'$ are pure. We choose ON terms
\[
\begin{array}{rcl}
t_{1} & = & x_{1}'\\
t_{2} & = & x_{1}x_{3}\\
t_{3} & = & x_{1}x_{3}'
\end{array}
\]
After assigning pure literals $x_{1}=x_{3}'=1$, the CNF set $C/t_{3}$ is given by
\[
\barr{rrrrrrrr}
1 & 2 & 3 & 4 & 5 & 6 & 7 & 8\\
0 & -1 & 0 & 1 & 0 & 0 & -1 & -1\\
0 & 0 & 0 & -1 & 0 & 0 & 0 & 1
\earr
\]
which is clearly satisfiable by $(1,0,0,0,d,d,d,d)$. Hence $C$ is satisfiable.

Thus in CNF satisfiability if we can assume that all pure literals have been assigned already or that there are no pure literals to be assigned. Similarly assume that all unit clauses $\{l\}$ have been eliminated by assignments, then the above corollary can be written analogously  for the CNF set with no pure literals and unit clauses as follows.

\begin{cor}\label{Cor:ONtermexpCNF}\emph{Let $C$ be a CNF set with no pure literals or unit clauses and let $T=\{t_{1},t_{2},\dots,t_{m}\}$ be an ON set of terms. Then the set $C$ is satisfiable iff there is an index $k$ in $[1,m]$ such that the system
\[
C/t_{k}
\]
is satisfiable. If $q$ is a satisfying assignment of this set, then $(q(t_{k})\|q)$ gives an assignment satisfying $C$. Every solution of a satisfiable CNF set $C$ arises in this form. 
}
\end{cor}

\subsection{ON expansion based algorithm for CNF-SAT}
Combining the above two cases we get a procedure for decomposition and solving satisfiability of a CNF set which is a generalization of the DPLL algorithm. This generalization is in respect of the splitting rule of DPLL which is defined by the ON functions $x,x'$ relative to a single variable $x$, while the decomposition considered in the following algorithm extends the notion of splitting relative to an ON set of terms in many variables.

Let $(C,X)$ denote the CNF set $C$ with the list of literals $X$. The algorithm described next does the task of decomposition of the SAT problem data $(C,X)$ to sub-problems by choosing an ON set of terms and distributes the sub-problems to independent nodes for computation of solutions. The parameter $n_{0}$ is the maximum number of variables $X$, below which the SAT problem is solved by direct search over all assignments.

\begin{algorithm}[Decomposition and Distribution]\emph{Decompose($(C,X)$)
\begin{enumerate}
\item \textbf{Input} $(C,X)$, $n_{0}$
\item \textbf{while} $|X|=n>n_{0}$, \textbf{repeat}
\item Choose an ON set of terms $T$ over a subset of \emph{literals} in $X$.
\item \emph{Decompose}\/: compute sub-problems 
\[
C_{i}=C/t_{i}\mbox{ for }t_{i}\in T
\]
compute resultant literals $X_{i}$ after partial assignment such that $t_{i}=1$.
\item \emph{distribute} $(C_{i},X_{i})$ to independent nodes for independent computation.
\end{enumerate}
}
\end{algorithm}

The above algorithm for distribution is then used in the main algorithm

\begin{algorithm}[Main]\emph{SolveSAT($(C,X)$)
\begin{enumerate}
\item \textbf{Input} $(C,X)$,$n_{0}$
\item \textbf{while} $|X|>n_{0}$, \textbf{repeat}
\item Determine all \emph{unit clauses} with subset of literals $X_{1}$. Assign $x=1$ for all $x\in X_{1}$. 
\[
(C,X)\leftarrow (C(x=1,x\in X_{1}),X-X_{1})
\]
\item Determine \emph{pure literals} $x_{1},\ldots,x_{r}$, make partial assignments $x_{i}=1$ for $i=1,\ldots,r$ denoted by $t_{r+1}=1$ as in (\ref{purelitONterms}) and denote $\tilde{X}$ the set of un-assigned literals 
\[
(C,X)\leftarrow (C/t_{r+1},\tilde{X})
\]
\item Decompose($(C,X)$)
\item At each independent node: \textbf{if} $|X|\leq n_{0}$ solve the satisfiability problem. \textbf{return} solution at the node \textbf{else} assign flag $SAT=F$ broadcast flag.
\end{enumerate}
}
\end{algorithm}

We illustrate the algorithm with an example.

\subsubsection{Example for CNF-SAT 2}
Consider the CNF set given by the matrix $C$, first row denoting column indices,
\[
C=
\barr{rrrrrrrr}
1 & 2 & 3 & 4 & 5 & 6 & 7 & 8 \\
1 & -1 & 0 & 0 & 1 & 0 & 0 &-1\\
-1 & 0 & 1 & -1 & 0 & 0 & 1 & 0\\
0 & 0 & -1 & 1 & 0 & 0 & 0 & 1\\
0 & 1 & 0 & 0 & -1 & 1 & -1 & 0\\
0 & 0 & 0 & 0 & 1 & -1 & -1 & 0
\earr
\]
and variables $X=\{x_{1},\ldots,x_{8}\}$ indexed by columns of $C$. There are no positive or pure literals. We choose ON set
\[
\{x_{1},x_{1}'x_{2},x_{1}'x_{2}'\}
\]
and decompose into
\[
C_{1}=C/x_{1}=
\barr{rrrrrrrr}
1 & 2 & 3 & 4 & 5 & 6 & 7 & 8\\
0 & 0 & 1 & -1 & 0 & 0 & 1 & 0\\
0 & 0 & -1 & 1 & 0 & 0 & 0 & 1\\
0 & 1 & 0 & 0 & -1 & 1 & -1 & 0\\
0 & 0 & 0 & 0 & 1 & -1 & -1 & 0
\earr
\]
\[
C_{2}=C/x_{1}'x_{2}=
\barr{rrrrrrrr}
1 & 2 & 3 & 4 & 5 & 6 & 7 & 8\\
0 & 0 & 0 & 0 & 1 & 0 & 0 & -1\\
0 & 0 & -1 & 1 & 0 & 0 & 0 & 1\\
0 & 0 & 0 & 0 & 1 & -1 & -1 & 0
\earr
\]
\[
C_{3}=C/x_{1}'x_{2}'
\barr{rrrrrrrr}
1 & 2 & 3 & 4 & 5 & 6 & 7 & 8 \\
0 & 0 & -1 & 1 & 0 & 0 & 0 & 1\\
0 & 0 & 0 & 0 & -1 & 1 & -1 & 0\\
0 & 0 & 0 & 0 & 1 & -1 & -1 & 0
\earr
\]
In $C_{1}$ pure literals are assigned as $x_{2}=x_{8}=1$. This reduces $C_{1}$ to
\[
\barr{rrrrrrrr}
1 & 2 & 3 & 4 & 5 & 6 & 7 & 8\\
0 & 0 & 1 & -1 & 0 & 0 & 1 & 0\\
0 & 0 & 0 & 0 & 1 & -1 & -1 & 0
\earr
\]
This leads to pure literal assignments $x_{3}=x_{5}=1$. Hence other variables $x_{4},x_{6},x_{7}$ can be assigned arbitrary. Hence $(C,X)$ is SAT even without checking the other cases $C_{2},C_{3}$.

\section{Conclusions}
This paper resolves the problem of solving Boolean systems of equations in many variables and $B_{0}$ co-efficients using general ON expansion without converting the system to single equation. The method also characterizes all solutions of such systems. As a central idea the ON expansion splits the original problem into a smaller sub-problems indexed by ON functions whose solutions need be searched only on the support of the specific ON function. Since the assignments in the support of an ON function are known apriori (or are easy to determine on the fly) the search of solution of the smaller problem is simplified. Further, since the solutions of the smaller problems are computed independently, this algorithm is inherently parallel. Algorithms for solving Boolean systems can be improved in their performance further by incorporating conjugate problems during computation. This aspect needs to be explored further.

The special case of expansion when the ON functions are ON terms is also shown to be useful in deciding satisfiability and characterization of solutions. This special procedure also leads to a generalization of the DPLL algorithm where the splitting stage is extended over many variables. The CNF-SAT algorithm developed using this special procedure is most valuable for decomposition of the problem for parallel computation. Methods developed in this paper, it is hoped, shall be useful for devising scalable parallel approaches of Boolean satisfiability problems by incorporating heuristics in constructing ON sets of functions.

\begin{center}
Acknowledgements

\noindent
Supported by the project grant 11SG010 of IRCC of IIT Bombay. Author gratefully acknowledges enlightening comments by Professor Rudeanu leading to improvements in the paper.
\end{center}


\begin{thebibliography}{xxxx}
\bibitem{bool} George Boole. An Investigation of the Laws of thought. Walton, London, 1854.
\bibitem{brow} F.\ M.\ Brown. Boolean reasoning. The logic of Boolean equations. Dover, 2006.
\bibitem{rude} Sergiu Rudeanu. Boolean functions and equations. North Holland, Amsterdam, 1974.
\bibitem{rud2} Sergiu Rudeanu. Lattice functions and equations. Springer Verlag, London, 2001.
\bibitem{hand} A.\ Biere, M.\ Heule, Hans van Maaren, T.\ Walsh (Eds). Handbook of Satisfiability. IOS Press, 2009.
\bibitem{crah} Yves Crama and Peter Hammer. Boolean functions. Theory, algorithms and applications. Encyclopedia of Mathematics and its applications, vol.142. Cambridge, 2011.
\bibitem{bard} Gregory Bard. Algebraic cryptanalysis. Springer 2009.
\bibitem{mezm} Marc Mezard and Andrea Montanari. Information, Physics and Computation. Oxford University Press, 2009.
\bibitem{hamw} Youssef Hammadi and C.\ M.\ Wintersteiger. Seven challenges in parallel SAT solving. Challenge paper AAAI 2012 Sub-Area spotlights track. Association of Advancement of Artificial Intelligence.
\bibitem{kohj} Kohavi and Jha, Switching and automata theory, Cambridge 2008.
\bibitem{sule} Generalization of Boole-Shannon expansion, consistency of Boolean equations and elimination by orthonormal expansion, arXiv.org:1306.2484v3,[cs.CC], December 6, 2013. 
\end{thebibliography}
\end{document}